\documentclass[
superscriptaddress,
amsmath,amssymb,
aps,
prl,
twocolumn,
reprint,
floatfix]{revtex4}
\usepackage{graphicx}
\usepackage{amssymb,amsmath,bbm,braket,indentfirst,bm}
\usepackage{epstopdf}
\usepackage{dcolumn}
\usepackage{color}
\usepackage{bigstrut}
\usepackage{empheq}
\usepackage{tcolorbox}
\usepackage{mathtools}

\def \g{\gamma}    \def \a{\alpha}
    \def \b{\beta} 
\def \s{\sigma}      
    
   \def \d{\delta} 
    \def \l{\lambda}

   \def \D{\Delta} 
 
\def \f{\frac} 

\def \>{\rangle} 
\def \<{\langle} 
 
\def\bra{\langle} 
\def \ket{\rangle}

\def\Rav{\bra R\ket}
\def\lav{\bra \l\ket}
\def\inv{^{-1}}
\def \Ke{K^\text{eff}}
\def \cil{\chi^{(\l)}}
\def \ciR{\chi^{(R)}}
\def \nul{\nu^{(\l)}}
\def \nuR{\nu^{(R)}}
\def \ol{\overline{\l}}
\def \oR{\overline{R}}
\def \sm{\setminus}

\newcommand{\be}{\begin{equation}}
\newcommand{\ee}{\end{equation}}
\newcommand{\bea}{\begin{eqnarray}}
\newcommand{\eea}{\end{eqnarray}}

\begin{document}

\title{Constrained optimization as ecological dynamics with applications to random quadratic programming in high dimensions}

\author{Pankaj Mehta}
\email{pankajm@bu.edu}
\affiliation{Physics Department, Boston University, Boston, Massachusetts 02215, USA}

\author{Wenping Cui}
\affiliation{Physics Department, Boston University, Boston, Massachusetts 02215, USA}
\affiliation{Physics Department, Boston College, Chesnut Hill, Massachusetts 02467, USA}

\author{Ching-Hao Wang}
\affiliation{Physics Department, Boston University, Boston, Massachusetts 02215, USA}

\author{Robert Marsland}
\affiliation{Physics Department, Boston University, Boston, Massachusetts 02215, USA}

\date{\today}

\begin{abstract}

Quadratic programming (QP) is a common and important constrained optimization problem. Here, we derive a surprising duality between constrained optimization with inequality constraints -- of which QP is a special case -- and consumer resource models describing ecological dynamics. Combining this duality with a recent `cavity solution', we analyze high-dimensional, random QP  where the optimization function and constraints are drawn randomly. Our theory shows remarkable agreement with numerics and points to a deep connection between optimization, dynamical systems, and ecology.
\end{abstract}

\maketitle

\newpage

Optimization is an important problem for numerous disciplines including physics, computer science, information theory, machine learning, and operations research \cite{boyd2004convex,bertsekas1999nonlinear, bishop:2006:PRML, mezard2009information}. Many optimization problems are amenable to analysis using techniques from the statistical physics  of disordered systems \cite{zdeborova2009statistical,mezard2002analytic, moore2011nature}. Over the last few years,  similar methods have been used to study community assembly and ecological dynamics suggesting a deep connection between ecological models of community assembly and optimization \cite{fisher2014transition, kessler2015generalized,  dickens2016analytically,  bunin2017ecological, advani2018statistical, barbier2018generic, biroli2018marginally, tikhonov2017collective, marsland2018available}.Yet, the exact relationship between these two fields remains unclear. 

Here, we show that constrained optimization problems with inequality constraints are naturally dual to an ecological dynamical system describing a generalized consumer resource model \cite{macarthur1967limiting,  macarthur1970species,chesson_macarthurs_1990}. As an illustration of this duality, we focus on a particular important and commonly encountered constrained optimization problem: Quadratic Programming (QP) \cite{boyd2004convex}. In QP, the goal is to minimize a quadratic objective function subject to inequality constraints. We show that QP is dual to one of the most famous models of ecological dynamics, MacArthur's Consumer Resource Model (MCRM) -- a system of ordinary differential equations describing how species compete for a pool of common resources \cite{macarthur1967limiting, macarthur1970species, chesson_macarthurs_1990}. We also show that the Lagrangian dual of QP has a natural description in terms of generalized Lotka-Volterra equations that can be derived from the MCRM in the limit of fast resource dynamics.

We then consider random quadratic programming (RQP) problems  where the optimization function and inequality constraints are drawn from a random distribution. We exploit a recent `cavity solution' to the MCRM by one of us to construct a mean-field theory for the statistical properties of RQP \cite{advani2018statistical}. Our theory is exact in infinite dimensions and shows remarkable agreement with numerical simulations even for moderately sized finite systems. This duality also allows us to use ideas from ecology to understand the behavior of RQP and interpret community assembly in the MCRM as an optimization problem. 
\section*{Optimization as ecological dynamics}
We begin by deriving the duality between constrained optimization and ecological dynamics. Consider an optimization problem of the form	 
\begin{equation}
\begin{aligned}
& \underset{\mathbf R}{\text{minimize}}
& & f({\mathbf R}) \\
& \text{subject to}
& & g_i({\mathbf R}) \leq 0, \; i = 1, \ldots, S.\\
&&& R_\alpha \geq 0, \; \alpha=1, \ldots, M.
\end{aligned}
\end{equation}
where the variables being optimized ${\mathbf R}=(R_1, R_2, \ldots, R_M)$ are constrained to be non-negative. We can introduce a  `generalized' Lagrange multiplier $\lambda_i$ for each of the $S$ inequality constraints in our optimization problem. In terms of the $\lambda_i$, we can write a set of conditions collectively known as the Karush-Kuhn-Tucker (KKT) conditions  that must be satisfied at any local optimum ${\mathbf R_{min}}$ of our problem \cite{boyd2004convex,bertsekas1999nonlinear, bishop:2006:PRML}. We note that for this reason, in the optimization literature the $\lambda_i$ are often called KKT-multipliers rather than Lagrange multipliers. The KKT conditions are:\\\\
\indent {\it Stationarity:} $\nabla_{\mathbf{R}}f({\mathbf R}_\mathrm{min}) +\sum_{j} \lambda_j \nabla_{\mathbf{R}} g_j({\mathbf R}_\mathrm{min})=0$\\
\indent {\it Primal feasibility:}  $g_i({\mathbf R}_\mathrm{min}) \leq 0$ \\
\indent {\it Dual feasibility:}  $\lambda_i \ge 0$ \\
\indent {\it Complementary slackness:} $\lambda_i(g_i({\mathbf R}_\mathrm{min})-m_i)=0$,\\\\
where the last three conditions must hold for all $i=1,\ldots,M$. The KKT conditions have a straightforward and intuitive explanation. At the optimum ${\mathbf R}_\mathrm{min}$, either $g_i({\mathbf R}_\mathrm{min})=0$ and the constraint is active $\lambda_i \ge 0$, or  $g_i({\mathbf R}_\mathrm{min}) \leq 0$ and the constraint is inactive  $\lambda_i=0$. In our problem, the KKT conditions must be supplemented with the additional requirement of positivity $R_\alpha \ge 0$.

\begin{figure}
\includegraphics[width=1.0\linewidth]{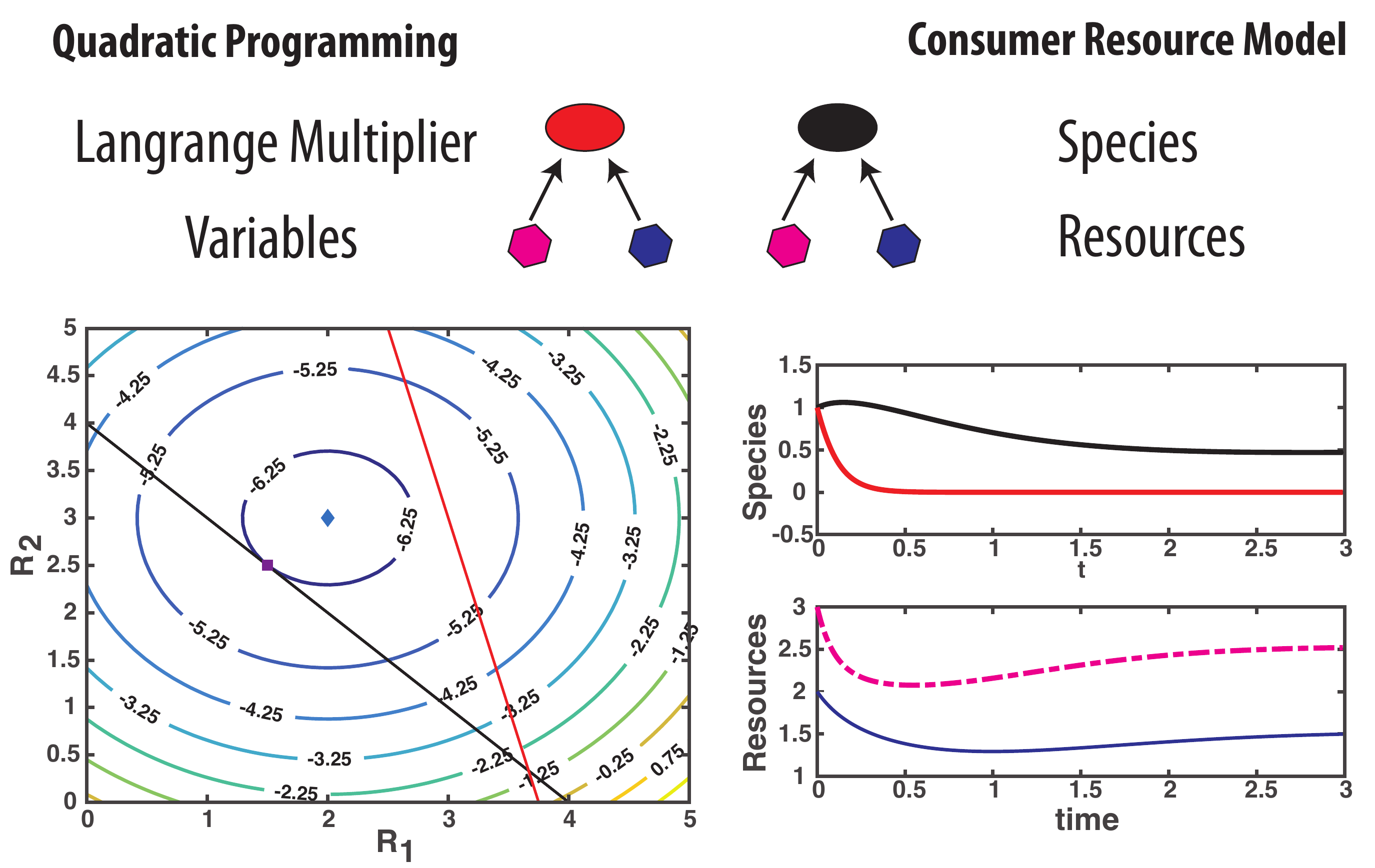}
\caption{{\bf Constrained optimization with inequality constraints is dual to an ecological dynamical system described by a generalized consumer
resource model (MCRM)}. The variables to be optimized (hexagons) and Lagrange multipliers (ovals) are mapped to resources
and species respectively. Species must consume resources to grow. (Bottom Left) A quadratic programming (QP) problem with two inequality constraints where the unconstrained optimum differs from the constrained optimum. (Bottom Right) Dynamics for MacArthur's Consumer Resource Model that is dual to this QP problem. The steady-state resource/species abundances correspond to the value of variables/Lagrange multipliers at the  QP optimum. For this reason, species corresponding to inactive constraints go extinct. }
\label{fig:figure1}
\end{figure}

One can easily show that the four KKT conditions and positivity are also satisfied by the steady states of the following set of differential equations restricted
to the space $\lambda_i, R_\alpha \ge 0$:
\bea
{d \lambda_i  \over dt}&=&\lambda_i g_i({\mathbf R}) \nonumber \\
{d R_{\alpha} \over dt} &=& [-\partial_{R_\alpha} f({\mathbf R}) -\sum_{j} \lambda_j \partial_{R_\alpha} g_j({\mathbf R})] R_\alpha
\label{GCRM1}
\eea
The first of these equations just describes exponential growth of a ``species'' $i$ with a resource-dependent ``growth rate'' $g_i({\mathbf R})$. Species with $g_i({\mathbf R}_\mathrm{min}) \leq 0$ correspond to constraints that are inactive and go extinct in the ecosystem (i.e $\lambda_{i \, \mathrm{min}} =0$), whereas species with $g_i({\mathbf R}_\mathrm{min})=0$ survive at steady state and correspond to active constraints with $\lambda_{i \, \mathrm{min}} \neq 0$ (see Figure \ref{fig:figure1} for a simple two-dimensional example). The second equation in (\ref{GCRM1}) performs a ``generalized gradient descent'' on the optimization function $f(\mathbf{R}) +\sum_j \lambda_j g_j(\mathbf{R})$ (note the extra factor of $R_\alpha$ in our dynamics compared to the usual gradient descent equations). In the context of ecology, these equations describe the dynamics of a set of resources $ \{ R_\alpha \}$ produced at a rate $-\partial_{R_\alpha} f({\mathbf R}) R_\alpha$ and consumed by individuals of species $j$ at a rate $ \lambda_j \partial_\alpha g_j({\mathbf R})R_\alpha$.

This suggests a simple dictionary for constructing systems dual to optimization problems with inequality constraints (see Figure \ref{fig:figure1}) . The variables are resources whose dynamics are governed by the gradient of the function being optimized. Each inequality is associated with a species through its corresponding Lagrange (KKT) multiplier. Species that survive in the ecosystem correspond to active constraints whereas species that go extinct correspond to inactive constraints. The steady-state values of the resource and species abundances correspond to the local optimum $\mathbf{R}_\mathrm{min}$ and Lagrange multipliers at the optimum $\{ \lambda_{j \mathrm{\,min}} \}$, respectively. Finally, the $f({\mathbf R}_\mathrm{min})$ are closely related to Lyapunov functions known to exist in the literature for specific choices of resource dynamics \cite{macarthur1970species, chesson_macarthurs_1990, tikhonov2017collective}.

\section*{Ecological duals of Quadratic Programming (QP)}

For the rest of the paper, we focus on QP where the optimization function is quadratic, $f(\mathbf{R})= {1 \over 2} \mathbf{R}^T Q \mathbf{R} + \mathbf{b}^T \mathbf{R}$, with $Q$ a positive semi-definite matrix, and linear inequality constraints. By going to the eigenbasis of $Q$, we can always rewrite the QP problem as minimizing a square distance
\begin{equation}
\begin{aligned}
& \underset{\mathbf R}{\text{minimize}}
& & \frac{1}{2} ||\mathbf{R}-\mathbf{K}||^2 \\
& \text{subject to}
& & \sum_{\alpha} c_{i \alpha}R_\alpha \leq m_i, \; i = 1, \ldots, S.\\
&&& R_\alpha \geq 0, \; \alpha=1, \ldots, M.
\label{QPeq}
\end{aligned}
\end{equation}
Using (\ref{GCRM1}), we can construct the dual ecological model:
\bea
{d \lambda_i  \over dt}&=&\lambda_i (\sum_\alpha c_{i \alpha} R_\alpha -m_i)  \nonumber \\
{d R_{\alpha} \over dt} &=& R_\alpha (K_\alpha -R_\alpha) -\sum_{j} \lambda_j c_{j \alpha} R_\alpha.
\label{GCRM2}
\eea
The is the famous MacArthur Consumer Resource Model (MCRM) which  was first introduced by Robert MacArthur and Richard Levins in their seminal papers \cite{macarthur_limiting_1967, macarthur1970species} and has played an extremely important role in theoretical ecology \cite{chesson2000mechanisms, tilman_resource_1982}.

In optimization problems, one often works with the Lagrangian dual of an optimization problem. We show in the appendix that the dual to (\ref{QPeq}) is just 
\begin{equation}
\begin{aligned}
& \underset{ \lambda_i}{\text{maximize}}
& & \sum_i  \lambda_i [\kappa_i -\frac{1}{2} \sum_j \alpha_{ij} \lambda_j]\\
& \text{subject to}
&& \lambda_i \geq 0,
\label{DualQPeq}
\end{aligned}
\end{equation}
with $\kappa_i = \sum_{\alpha} K_\alpha (c_{i \alpha}-m_i)$, $\alpha_{ij}=\sum_{\alpha}c_{i \alpha}c_{j \alpha}$, and the sum restricted to $\alpha$ for which $R_{\alpha \mathrm{\, min}} \neq 0$. It is once again straightforward to check that the local minima of this problem are in one-to-one correspondence with steady states of the Generalized Lotka-Volterra Equations (GLVs) of the form:
\be
{d \lambda_i \over dt}= \lambda_i (\kappa_i -\sum_j \alpha_{ij} \lambda_j)
\label{GLV}
\ee
As with the primal problem, the species in the GLV have a natural interpretation as Lagrange multipliers enforcing inequality constraints. This GLV can also be directly obtained from the MCRM in (\ref{GCRM2})  in the limit where the resource dynamics are extremely fast by setting ${dR_\alpha \over dt}=0$ in the second equation  and plugging in the steady-state resource abundances into the first equation  \cite{macarthur1970species, chesson_macarthurs_1990} (see Appendix).  This shows the  Lagrangian dual of QP maps to a dynamical system described by a GLV -- which itself can be derived from the MCRM which is the dynamical dual to the primal optimization problem!  

\section*{Random Quadratic Programming (RQP)}
Recently, the MCRM was analyzed in the high-dimensional limit where the number of resources and species in the regional species pool is large ($S,M \gg 1$). In this limit, the resource dynamics were extremely complex, with many resources deviating significantly from their unperturbed values and a large fraction of species in the regional pool going extinct  \cite{advani2018statistical}. In terms of the corresponding optimization problem, this suggests that $f({\mathbf R}_\mathrm{min})$ will generically be far from zero and many of constraints will be inactive.

\begin{figure}[t]
\includegraphics[width=1.0\linewidth]{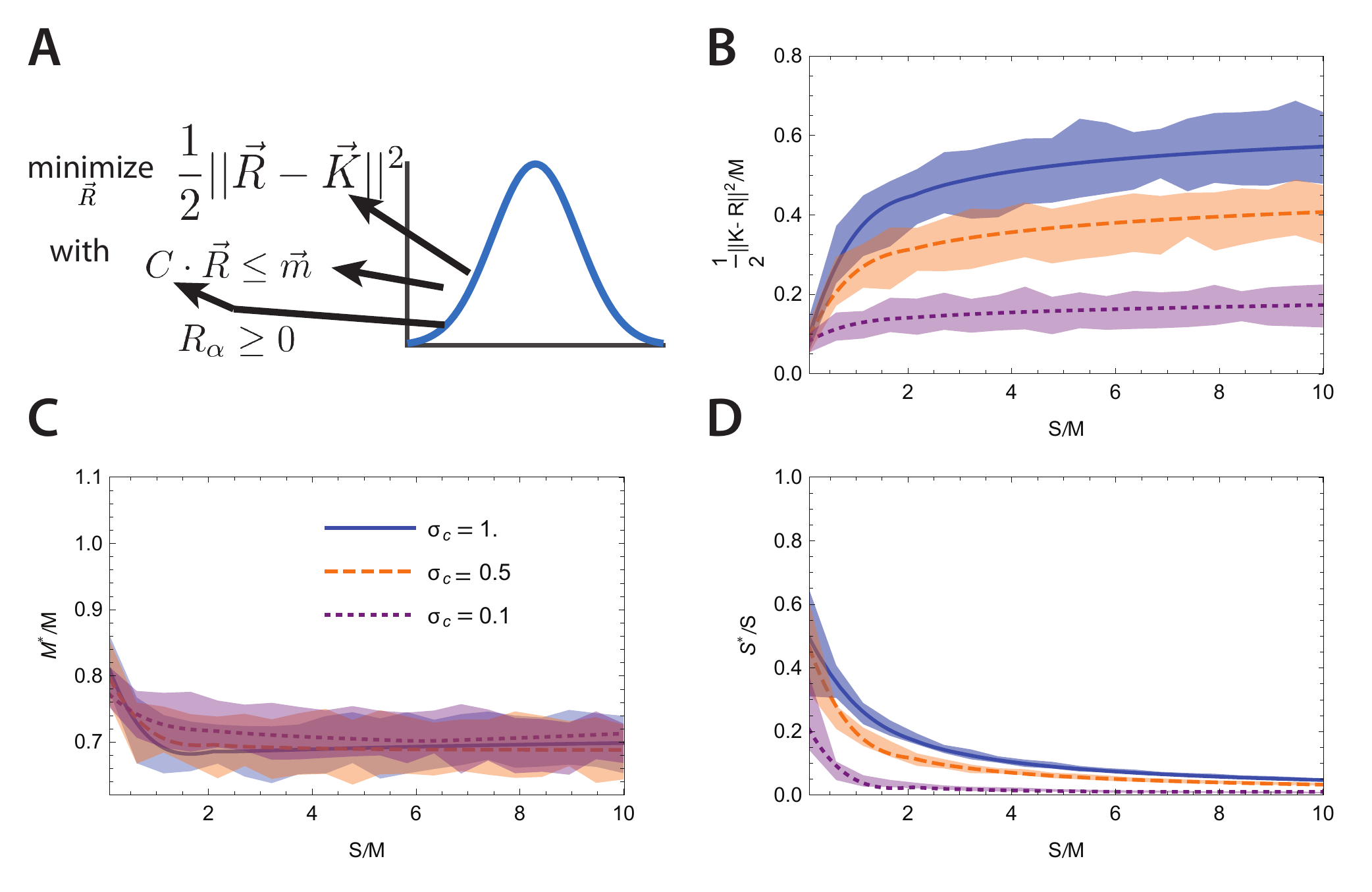}
\caption{{\bf Random Quadratic Programming (RQP)}. (A) In RQP, the parameters of the quadratic optimization function and inequality constraints are drawn from
a random distribution. Effect of varying the ratio of constraints to variables $S/M$ on (B) the value of the optimization function $f({\mathbf R}_\mathrm{min})/M$, (C) the fraction of non-zero variables $\frac{M^*}{M}$ and (D) the fraction of active constraints $\frac{S^*}{S}$. Cavity solutions are solid lines and shaded region show $\pm1$ standard deviation from 50 independent optimizations of RQP using the CVXOPT  package in Python 3 with $M=100$, $\mu_c=1$, $K=1$, $\sigma_K=1$, $m=1$, $\sigma_m=0.1$. Code is available
in supplementary files. }
\label{fig:fig2}
\end{figure}

To better understand this, we analyzed Random quadratic programming (RQP) problems in high dimension. In RQP,  the parameters in (\ref{QPeq}) are drawn from random distributions (see Figure \ref{fig:fig2}A). We focus on the case where the $K_\alpha$ and $m_i$ are independent random normal variables drawn from Gaussians with means $K$ and $m$ and variances $\sigma_K^2$ and $\sigma_m^2$, respectively. The elements of the constraint matrix $c_{i \alpha}$ are also drawn from Gaussians with mean $\mu_c/M$ and variance $\sigma_c^2/M$ \footnote{We note that this scaling is slightly different from that in \cite{advani2018statistical} where the elements where chosen to scale with $S$ not $M$. This choice does not change the results, but results in slightly different expressions}.This scaling with $M$ is necessary to ensure that the sum that appears in the inequality constraints in  (\ref{QPeq}) has a good thermodynamic limit when $M,S \rightarrow \infty$ with $M/S=\gamma$ held fixed.  

We are especially interested in understanding the statistical properties of solutions to the RQP (see Fig. \ref{fig:fig2}A) . Among the quantities we examine are the expectation value of the optimized function at the minima $ \langle f({\mathbf R}_\mathrm{min})\rangle/M$, the fraction of active constraints, $S^*/S$, the fraction of variables that are non-zero at the optimum, $M^*/M$, as well the first two moments of $R_{\alpha \mathrm{min}}$ and $\lambda_{j \min}$ (see Appendix for details).

It is possible to a derive mean-field theory (MFT) for the statistical properties of the optimal solution in the RQP -- or correspondingly the steady-states of the MCRM -- using the cavity method. The basic idea behind the cavity method is to derive self-consistency equations that relate the optimization problem (ecosystem) with $M+1$ variables (resources) and $S+1$ inequality constraints (species) to a problem where a constraint (species) and variable (resource) have been removed: $(M+1, S+1) \rightarrow (M,S)$ \cite{advani2018statistical}. The need to remove both a constraint and variable is important for keeping all order one terms in the thermodynamic limit \cite{mezard1989space, ramezanali2015critical}. In what follows, we focus on the replica-symmetric solution.

The cavity equation exploits the observations the constraint  $\sum_{\alpha=1}^M c_{i \alpha} R_{\alpha}$  is a sum of many  random variables, $c_{i \alpha}$. When $M \gg 1$, due to the law of large numbers we can model such a sum by a random variable drawn from a Gaussian whose mean and variance involve the statistical quantities described above. Less obvious from the perspective of QP is that we need to introduce a second mean-field quantity $K_\alpha^{eff}$ (see Appendix and \cite{advani2018statistical}).  After introducing the Lagrange multipliers that enforce the inequality constraints, the optimization function to be minimized takes the form
\begin{align}
\frac{1}{2} ||\mathbf{R}&-\mathbf{K}||^2 + \sum_j \lambda_j (c_{j \alpha}R_\alpha-m_j)\nonumber\\
&=  {1 \over 2} \sum_{\alpha} \left\{ R_\alpha [R_\alpha-K_\alpha^{eff}(\lambda)] +  K_\alpha [K_\alpha - R_\alpha]\right\}  \nonumber,
\end{align}
where we have defined the mean-field variable
$$
K_\alpha^{eff}(\lambda) = K_\alpha -\sum_{j=1}^S \lambda_j c_{j \alpha}.
$$
Since $K_\alpha^{eff}(\lambda)$ is also a sum of many terms containing $c_{i \alpha}$, it can also be approximated as a random variable drawn from a Gaussian whose mean and variance are calculated self-consistently .

The full derivation of the replica symmetric mean-field equations is identical to that in  \cite{advani2018statistical} and is given in the Appendix. The resulting self-consistent mean-field cavity equations can be solved numerically in Mathematica. Figure \ref{fig:fig2} shows the results of our mean-field equations and comparisons to numerics where we directly optimize the RQP problem over many independent realizations using the CVXOPT package in Python \cite{andersen2013cvxopt} . Notice the remarkable agreement between our MFT and results from direct optimization even for moderate system sizes with $M=100$.  In the Appendix, we show that the cavity solution can also accurately describe the dual MCRM.

 Figure \ref{fig:fig2} also shows that the statistical properties of the QP solutions change as we vary the number of constraints $S$ and the variance of the constraint matrix $c_{i \alpha}$. When $S\ll M$, the expectation value of the optimization function  $f({\mathbf R}_\mathrm{min})/M$ approaches zero -- the minimum for the unconstrained problem. In this limit, the few constraints that are present are also active. As  $S/M$ is increased, the fraction of active constraints quickly drops, $f({\mathbf R}_\mathrm{min})/M$ quickly increases, after which both quantities reach a plateau where they vary very slowly with $S$. The value of the the plateau depends on $\sigma_c$. Increasing the variance of the constraints results in more active constraints and a larger value of  $f({\mathbf R}_\mathrm{min})$ at the optimum. 
 
These results about RQP can be naturally understood using ideas from ecology. Intuitively, a smaller $\sigma_c$ means more ``redundant'' constraints. In ecology, this is the principle of limiting similarity: species with large niche overlaps (similar $c_{i \alpha}$ ) competitively exclude each other \cite{macarthur_limiting_1967, macarthur1970species,  chesson_macarthurs_1990, chesson2000mechanisms, tilman_resource_1982}.  In the language of optimization, this ecological intuition suggests that when constraints are similar enough, only the most stringent of these will be active due to an effective competitive exclusion between constraints. Thus, in RQP competitive exclusion becomes a statement about the geometry of how random planes in high dimension repel each other at the corners of simplices. In all cases, increasing $S$ increases the total number of active constraints (species) even though the fraction of active constraints decreases. For this reason, the optimization problem is more constrained for larger $S$ and $f({\mathbf R}_\mathrm{min})/M$ is larger. Finally the plateau in statistical quantities at large $S$ can be understood as arising from what in ecology has been called ``species packing'' -- there is a capacity to the number of distinct species that any ecosystem can typically support \cite{macarthur_limiting_1967, macarthur1970species}. 

\section*{Discussion}

In this paper, we have derived a surprising duality between constrained optimization problems and ecologically inspired dynamical systems. We showed that QP (in any dimension) maps to one of the most famous models of ecological dynamics, MacArthur's Consumer Resource Model (MCRM) -- a system of ordinary differential equations describing how species compete for a pool of common resources. By combining this mapping with a recent  `cavity solution' to the MCRM, we constructed a mean-field theory for the statistical properties of RQP that showed remarkable agreement with numerical simulations. Intuitions from ecology suggest that the geometry of constrained optimization can be described using a competitive exclusion between constraints which in our case correspond to random high-dimensional hyperplanes. This work suggests that  the deep connection between geometry, ecology, and high-dimensional random ecosystems is a generic property of a large class of  generalized consumer resource models \cite{landmann2018phase}. Our works also gives a natural explanation of the existence of Lyapunov functions in these models.

\section{Acknowledgments}
The work was supported by NIH NIGMS grant 1R35GM119461,  Simons Investigator in the Mathematical Modeling of Living Systems (MMLS) to PM, and the Scialog Program sponsored jointly by Research Corporation for Science Advancement (RCSA) and the Gordon and Betty Moore Foundation.

\bibliography{refs_ecology_QP.bib}

\onecolumngrid
\newpage
\appendix
\section{ Derivation of Lagrangian dual for QP}
In this section, we derive the Lagrangian dual to our primal Quadratic Programming (QP) problem
\begin{equation}
\begin{aligned}
& \underset{\mathbf R}{\text{minimize}}
& & \frac{1}{2} ||\mathbf{R}-\mathbf{K}||^2 \\
& \text{subject to}
& & \sum_{\alpha} c_{i \alpha}R_\alpha \leq m_i, \; i = 1, \ldots, S.\\
&&& R_\alpha \geq 0, \; \alpha=1, \ldots, M.
\label{QPeq2}
\end{aligned}
\end{equation}
We start by introducing Lagrange (KKT) multipliers $\lambda_i$ dual to each of the $S$ constraints and Langrange KKT (multipliers) $\mu_\alpha$. that enforce positivity. Then, the function to be optimized is
\be
\begin{aligned}
& \underset{\lambda_j}{\text{maximize}}
& \underset{{\mathbf R}_\alpha}{\text{minimize}}
&\,  \frac{1}{2} \sum_{\alpha} (R_\alpha^2 -2 K_\alpha R_\alpha + K_\alpha^2) +\sum_{j,\alpha} \lambda_j (c_{j \alpha} R_\alpha -m_i) -\mu_\alpha R_\alpha \\
& \text{subject to}
&& \lambda_j \geq 0\; j=1, \ldots, S
\label{QPeqDual}
\end{aligned}
\end{equation}
We take the derivative with respect to $R_\alpha$ and note that 
\be
R_{\alpha *} = \mathrm{max}[0, K_\alpha -\sum_j c_{j \alpha} \lambda_j]
\ee
where we have used the KKT condition $\mu_\alpha R_{\alpha*}=0$

Plugging this back into (\ref{QPeqDual}), we find that the function to be maximized with respect to the $\lambda_i$ is
\be
\sum_i \lambda_i [ \kappa_i -\frac{1}{2} \sum_j \alpha_{ij} \lambda_j]\
\ee
with 
\be
\kappa_i = \sum_{\alpha, R_{\alpha*} \neq 0} K_\alpha c_{i \alpha}  -m_i
\ee
 and 
 \be
 \alpha_{ij}=\sum_{\alpha, R_{\alpha*} \neq 0}c_{i \alpha}c_{j \alpha}.
\ee
\section{ Derivation of Lotka Volterra Equations form MCRM}
We start from the MCRM dynamical equations
\bea
{d \lambda_i  \over dt}&=&\lambda_i (\sum_\alpha c_{i \alpha} R_\alpha -m_i)  \nonumber \\
{d R_{\alpha} \over dt} &=& R_\alpha [(K_\alpha -R_\alpha) -\sum_{j} \lambda_j c_{j \alpha} ]R_\alpha.
\label{GCRM3}
\eea
Notice that setting the second equation to zero we get
\be
R_{\alpha *} = \mathrm{max}[0, K_\alpha -\sum_j c_{j \alpha} \lambda_j].
\ee
Plugging this into the first equation in (\ref{GCRM3}) gives
\be
{d \lambda_i \over dt}= \lambda_i (\kappa_i -\sum_j \alpha_{ij} \lambda_j)
\ee
with $\alpha_{ij}$ and $\kappa_i$ defined as in the last appendix.
\section{Additional figure comparing RQP, MCRM, and MFT}
In this section, we supplement Figure 2 in main text  with an additional figure showing a comparison of  the Cavity solution, optimization of RQP, and steady-state values of the MCRM dual to the RQP. For each choice of parameters, the RQP were solved using the CVXOPT package in Python 3. The dual MCRM was constructed as outlined in main text and then integrated to steady-state using standard ODE solvers in Python. See supplementar
\begin{figure}[h]
\includegraphics[width=1.0\linewidth]{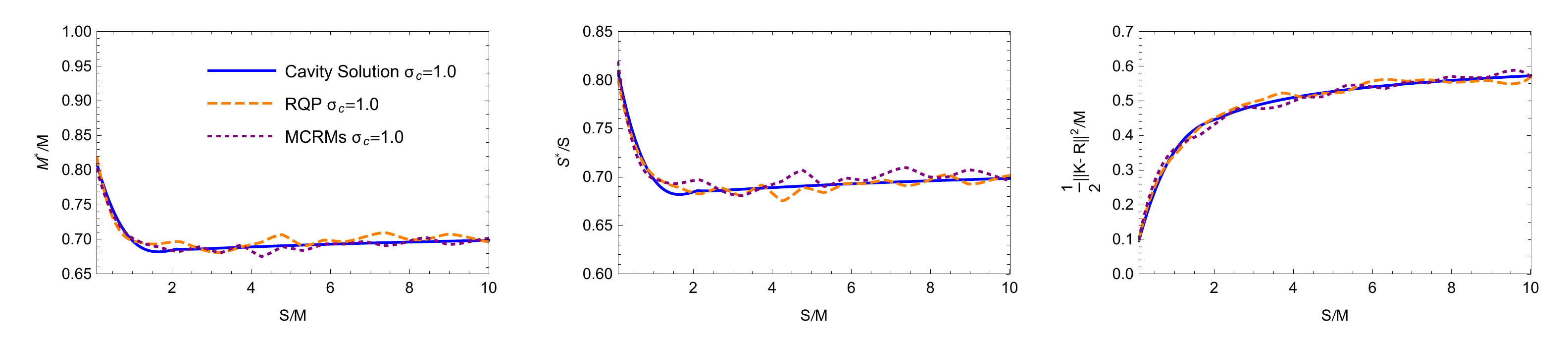}
\caption{{\bf Comparison of Cavity Solution (solid line), RQP (long dash line), and dual MCRMs (short dash line)}. The simulations represent averages from 50 independent
realizations and parameters as in Figure 2 of main text.}
\label{fig:figSI}
\end{figure}

\section{ Derivation of cavity solution}

\subsection{Model setup}
In this section, we derive the cavity solution to the MCRM (Eq. \eqref{GCRM2} in the main text)
\bea
{d \lambda_i  \over dt}&=&\lambda_i \left(\sum_\alpha c_{i \alpha} R_\alpha -m_i\right)  \nonumber \\
{d R_{\alpha} \over dt} &=& R_\alpha (K_\alpha -R_\alpha) -\sum_{j} \lambda_j c_{j \alpha} R_\alpha.
\label{GCRM}
\eea
Note that here we follow closely the derivation in \cite{advani2018statistical}. The only difference is that here we consider the consumer preference $c_{i\alpha}$ as random variables drawn from a Gaussian distribution with mean $\mu_c/M$ and variance $\sigma_c^2/M$, as opposed to the choices $\mu_c/S$ and $\sigma_c^2/S$ used in that work. With these definitions, we can decompose the consumer preference into $c_{i\alpha}=\mu_c/M + \sigma_c d_{i\alpha}$, where the fluctuating part $d_{i\alpha}$ obeys
\bea \label{eq:d-mean}
\bra d_{i\a} \ket &=&0\\ \label{eq:d-var}
\bra d_{i\b} d_{j\b} \ket &=&\frac{\d_{ij}\d_{\a\b}}{M}.
\eea
We also assume that both the carrying capacity $K_\a$ and the minimum maintenance cost $m_i$ are independent Gaussian random variables with mean and covariance given by 
\bea
\bra K_\a\ket &=& K\\
\text{Cov} (K_\a, K_\b) &=& \d_{\a\b}\s^2_K\\
\bra m_i\ket &=& m\\
\text{Cov} (m_i, m_j) &=& \d_{ij}\s^2_m
\eea
Let $\bra R\ket =(1/M)\sum_{\a}R_{\a}$ and $\bra \l\ket =(1/S)\sum_{i}\l_i$ be the average resource and average species abundance, respectively. With all these defined, we can re-write Eq. \eqref{GCRM} as
\bea\label{eq:MCRMfluc-l}
\f{d\l_i}{dt} &=&\l_i\left\{[\mu_c\Rav - m] +\s_c\sum_\a d_{i\a}R_\a -\d m_i\right\}\\\label{eq:MCRMfluc-R}
\f{dR_\a}{dt} &=& R_\a\left\{[K-\mu_c\g\inv\lav]- R_\a -\s_c\sum_j d_{j\a}\l_j +\d K_\a\right\},
\eea
where $\d K_\a = K_\a -K, \d m_i = m_i-m$ and $\lambda = M/S$. We can interpret the bracketed terms in these equations as population mean growth rate and effective resource capacity, respectively, \emph{viz.}
\begin{empheq}[box=\fbox]{align}
\label{eq:g-def}
g &\equiv  \mu_c\Rav - m\\ \label{eq:Ke-def}
\Ke &\equiv  K-\mu_c\g^{-1} \bra \l \ket.
\end{empheq}

As noted in the main text, the basic idea of cavity method is to relate an ecosystem with $M+1$ resources (variables) and $S+1$ species (inequality constraints) to that with $M$ resources and $S$ species. Following Eq.\eqref{eq:MCRMfluc-l}\eqref{eq:MCRMfluc-R}, one can write down the ecological model for the $(M+1, S+1)$ system where resource $R_0$ and species $\l_0$ are introduced to the $(M,S)$ system as: 
\bea
\f{d\l_i}{dt} &=&\l_i\left\{g +\s_c\sum_\a d_{i\a}R_\a +\s_c d_{i0}R_0-\d m_i\right\}\label{lambdanew}\\
\f{dR_\a}{dt} &=& R_\a\left\{\Ke- R_\a -\s_c\sum_j d_{j\a}\l_j -\s_cd_{0\a}\l_0 +\d K_\a\right\},\label{rnew}
\eea
where all sums from now on are understood to be over the indices $\alpha, j > 0$ from the $(M,S)$ system. The equations for the newly introduced species ($i=0$) and resource $(\a =0)$ are given by 
\bea\label{eq:cav-l}
\f{d\l_0}{dt} &=&\l_0\left\{g +\s_c\sum_\a d_{0\a}R_\a + \s_c d_{00}R_0-\d m_0\right\}\\ \label{eq:cav-R}
\f{dR_0}{dt} &=& R_0\left\{\Ke- R_0 -\s_c\sum_j d_{j0}\l_j -\s_c d_{00}\l_0+\d K_0\right\},
\eea

\subsection{Deriving the self-consistency equations with cavity method}
Following the same procedure in \cite{advani2018statistical}, we introduce the following susceptibilities:
\be
\cil_{i\b}=\f{\partial \ol_i}{\partial K_\b}
\ee
\be
\ciR_{\a\b}=\f{\partial \oR_\a}{\partial K_\b}
\ee
\be
\nul_{ij}=\f{\partial \ol_i}{\partial m_j}
\ee
\be
\nuR_{\a j}=\f{\partial \oR_\a}{\partial m_j},
\ee
where we denote $\overline{X}$ as the steady-state value of $X$. Recall that the goal is to derive a set of self-consistency equations that relates the ecological system (optimization problem) characterized by $M+1$ resources (variables) and $S+1$ species (constraints) to that with the new species and new resources removed: $(S+1, M+1)\rightarrow (S,M)$. To simplify notation, denote $\overline{X}_{\sm 0}$ be the steady-state value of quantity $X$ in the absence of the new resource and new species. Since the introduction of a new species and resource represents only a small (order $1/M$) perturbation to the original ecological system, we can express the steady-state species and resource abundances in the $(S+1, M+1)$ system with a first-order Taylor expansion around the $(S,M)$ values. We note that the new terms $\sigma_c d_{i0}R_0$ in Eq. \eqref{lambdanew} and $\sigma_c d_{0\alpha}\lambda_0$ in Eq. \eqref{rnew} can be treated as perturbations to $m_i$, and $K_\alpha$, respectively, yielding:
\be\label{eq:lcav}
\ol_i =\ol_{i\sm 0}-\s_c\sum_\b\cil_{i\b}d_{0\b}\ol_0 -\s_c\sum_j\nul_{ij}d_{j0}\oR_0
\ee 
\be\label{eq:Rcav}
\oR_\a  =\oR_{\a\sm 0}-\s_c\sum_\b\ciR_{\a\b}d_{0\b}\ol_0-\s_c\sum_j\nuR_{\a j}d_{j0}\oR_0. 
\ee
The next step is to plug Eq.\eqref{eq:lcav}\eqref{eq:Rcav} into Eq.\eqref{eq:cav-l}\eqref{eq:cav-R} and solve for the steady-state value of $\l_0$ and $R_0$.

For the new species, setting Eq.\eqref{eq:cav-l} to zero and plugging in Eq.\eqref{eq:Rcav} gives
\be
0=\ol_0\left[g +\s_c\sum_\a d_{0\a}\oR_{\a\sm 0}-\s_c^2 \sum_{\a\b}\ciR_{\a\b}d_{0\a}d_{0\b}\ol_0-\s_c^2\sum_{\a j}\nuR_{\a j}d_{0\a}d_{j0}\oR_0 - \d m_0 +\s_c d_{00}\oR_0 \right].
\ee
We now note that each of the sums in this equation is the sum over a large number of uncorrelated random variables, and can therefore be well approximated by Gaussian random variables for large enough $M$ and $S$. It is a straightforward exercise to show that the mean and variance of the third sum as well as the variance of the second sum are all order $1/M$ or higher, and can be ignored in comparison to the order 1 terms. The mean of the second sum is
\begin{align}
\sum_{\alpha\beta} \langle \chi_{\alpha\beta}^{(R)}\rangle\langle d_{0\alpha}d_{0\beta}\rangle = \frac{1}{M} \sum_{\alpha}\langle \chi_{\alpha\alpha}^{(R)} \rangle = \chi 
\end{align}
where we have used the statistics of $d_{i\alpha}$ as defined in Eqs. \eqref{eq:d-mean}\eqref{eq:d-var}, and have defined $\chi \equiv \langle \chi_{\alpha\alpha}^{(R)}\rangle$. 

Using these observations about the second and third sums, we obtain
\bea\label{eq:cav-l0}
0&=& \ol_0\left[g -\s_c^2 \chi \ol_0 +\s_c\sum_\a d_{0\a}\oR_{\a\sm 0} -\d m_0  \right] + \mathcal{O}(M^{-1/2}),
\eea
Since the $m_i$ come from a Gaussian distribution, we can model the combination of the remaining sum with $\delta m_i$ by a single Gaussian random variable with zero mean and variance $\s_g^2$ given by
\bea
\s_g^2 &\equiv& \text{Var} \left(\s_c\sum_\a d_{0\a}\oR_{\a\sm 0} -\d m_0\right )\\
&=& \text{Var} \left(\s_c\sum_\a d_{0\a}\oR_{\a\sm 0}\right)+ \text{Var} \left(\d m_0\right)\\
&=& \s_c^2 \f{1}{M}\sum_\a\oR_{\a\sm 0}^2+ \s_m^2\\
&=& \boxed{\s_c^2 q_R + \s_m^2},
\eea
where
\be
q_R = \f{1}{M}\sum_\a \oR^2_{\a\sm 0}.
\ee 
Denoting $z_\l$ as a random variable with zero mean and unit variance, we can express Eq.\eqref{eq:cav-l0} in terms of the quantities just defined:
\be
0=\ol_0\left(g -\s_c^2\chi\ol_0 +\s_gz_\l\right).
\ee
Inverting this equation one gets
\be\label{eq:cav-l0-final}
\boxed{\ol_0 =\f{\text{max} [0, g+\s_g z_\l]}{\s_c^2\chi}}\,,
\ee
which is a truncated Gaussian. 

We can follow the same procedure to solve for the steady state of the resource. Setting Eq.\eqref{eq:cav-R} to zero and plugging in Eq.\eqref{eq:lcav} gives 
\be
0 = \oR_0\left(\Ke -\oR_0 -\s_c\sum_j d_{j0}\ol_{j\sm 0} + \s_c^2\sum_{j\b}\cil_{i\b}d_{j0}d_{0\b}\ol_0 + \s_c^2\sum_{jk}\nul_{jk}d_{j0}d_{k0}\oR_0+\d K_0-\s_c d_{00}\ol_0\right).
\ee 
Keeping only the leading order terms one arrives at
\be\label{eq:cav-R0}
0\approx \oR_0\left(\Ke -\oR_0+\d K_0 -\s_c\sum_j d_{j0}\ol_{j\sm 0} +\s_c^2\gamma^{-1} \nu R_0\right).
\ee
where $\nu \equiv \langle \nul_{jj}\rangle$ is the average susceptibility. 
As before, $\d K_0 -\s_c\sum_j d_{j0}\ol_{j\sm 0}$ is a Gaussian random variable with zero mean and variance $\s_{\Ke}^2$ given by
\bea
\s_{\Ke}^2 &\equiv& \text{Var} \left(\d K_0 -\s_c\sum_j d_{j0}\ol_{j\sm 0}\right )\\
&=& \text{Var} \left(\d K_0\right)+ \text{Var} \left(\s_c\sum_j d_{j0}\ol_{j\sm 0}\right)\\
&=& \s_K^2 +\s_c^2\f{1}{M}\sum_j\ol_{j\sm 0}^2 \\
&=& \boxed{\s_K^2 + \s_c^2 \g^{-1} q_\l} ,
\eea
where
\be
q_\l = \f{1}{S}\sum_j \ol^2_{j\sm 0}.
\ee
Denoting $z_{R}$ as a random variable with zero mean and unit variance, we can express Eq.\eqref{eq:cav-R0} in terms of the quantities just defined:
\be
0 = \oR_0\left(\Ke -\oR_0 +\s_{\Ke}z_R +\s_c^2\g^{-1}\nu\oR_0\right).
\ee
Finally, inverting this equation gives the steady-state distribution of the resource
\be\label{eq:cav-R0-final}
\boxed{\oR_0=\f{\text{max} (0, \Ke +\s_{\Ke}z_R)}{1-\g^{-1}\s_c^2\nu}}
\ee
 
 
Next let's examine the self-consistency equations for the fraction of non-zero species and resources, $\phi_\l$ and $\phi_R$, respectively. Note that the goal is to find the values of $\{\phi_\l, \phi_R, \bra \l\ket, \bra R\ket, q_R, q_\l,\chi,\nu \}$ with given sets of parameters $\{K, \s_K, m,\s_m, \mu_c, S, M\}$. By variable counting, we'll need eight equations to solve for these eight unknowns but so far we've only got two, Eq.\eqref{eq:cav-l0-final} and Eq.\eqref{eq:cav-R0-final}. To find the remaining six equations, let's define some quantities (c.f. Eq.\eqref{eq:g-def}\eqref{eq:Ke-def}):
\begin{empheq}[box=\fbox]{align}
\D_g&\equiv \f{g}{\s_g}= \f{\mu_c\bra R\ket -m}{\s_g}\\
\D_{\Ke}&\equiv \f{\Ke}{\s_{\Ke}} =\f{K-\mu_c\g^{-1}\bra\l\ket}{\s_{\Ke}},
\end{empheq}
as well as the function
\begin{empheq}[box=\fbox]{align}\label{eq:w-def}
w_j(\D)=\int_{-\D}^\infty \frac{dz}{\sqrt{2\pi}}e^{-\f{z^2}{2}}(z+\D)^j,
\end{empheq}
which will simplify our notation later.
First let's derive the self-consistency equation for the susceptibilities. This is done by taking the derivative of Eq.\eqref{eq:cav-R0-final} with respect to $K$ and of Eq.\eqref{eq:cav-l0-final} with respect to $m$ while noting the definition of $\phi_\l$ and $\phi_R$:
\begin{empheq}[box=\fbox]{align}
\nu &= -\f{\phi_\l}{\s_c^2\chi}\\
\chi &= \f{\phi_R}{1-\g^{-1}\s_c^2\nu}.
\end{empheq}
Since Eq.\eqref{eq:cav-l0-final} and Eq.\eqref{eq:cav-R0-final} imply that the species and resource distributions are truncated Gaussians, it will be useful to note the following:
\begin{tcolorbox}
Let $y=\text{max}\left(0, \f{a}{b}+\f{c}{b}z\right)$, with $z$ being a Gaussian random variable with zero mean and unit variance. Then its $j$-th moment is given by
\be
\bra y^j\ket = \left(\f{b}{c}\right)^j \int_{-\f{b}{a}}^\infty \frac{dz}{\sqrt{2\pi}}e^{-\f{z^2}{2}}\left( z+\f{b}{a}\right)^j.
\ee
\end{tcolorbox}

With this we can easily write down the self-consistency equations for the fraction of non-zero species and resources as well as the moments of their abundances (c.f. Eq.\eqref{eq:cav-l0-final} and Eq.\eqref{eq:cav-R0-final}): 
\begin{empheq}[box=\fbox]{align}
\phi_\l &=w_0(\D_g)\\
\phi_R &=w_0(\D_{\Ke})\\
\bra \l\ket &=\f{\s_g}{\s_c^2 \chi}w_1(\D_g)\\
\bra R\ket &=\f{\s_{\Ke}}{1-\g^{-1}\s_c^2 \nu}w_1(\D_{\Ke})\\
q_\l&=\bra \l^2\ket =\left(\f{\s_g}{\s_c^2 \chi}\right)^2w_2(\D_g)\\ 
q_r&=\bra R^2\ket =\left(\f{\s_{\Ke}}{1-\g^{-1}\s_c^2 \nu}\right)^2w_2(\D_{\Ke})\label{eq:cav-R-2moment}.
\end{empheq}
Note that we only write down the first and the second moments since these six equations, along with Eq.\eqref{eq:cav-l0-final} and Eq.\eqref{eq:cav-R0-final}, complete the equations required to solve for the eight variables. 

\subsection{Cavity solution to the optimization function}

Here we derive the cavity solution to the optimization function $f(\mathbf{R})$ defined as 
\bea
\bra f(\mathbf{R})\ket &=&\frac{1}{2} \bra||\mathbf{R}-\mathbf{K}||^2\ket \\
& =&\f{1}{2} \sum_{\alpha} \bra R_\alpha^2\ket -2 \bra K_\alpha R_\alpha\ket + \bra K_\alpha^2\ket .
\eea
The first term is given by Eq.\eqref{eq:cav-R-2moment} while the last term is just $K^2+\s_K^2$. What remains to be solved is $\langle K_\a R_\a\rangle$. From Eq.\eqref{eq:cav-R0-final}, one can write
\be
R_\a (K_\a)=\f{\text{max} (0, K_\a - \mu_c\g^{-1}\bra \l\ket + z_\lambda \sqrt{\s_c ^2\g^{-1} q_\l})}{1-\g^{-1}\s_c^2\nu}.
\ee
Now let variable $k$ be drawn from the same distribution as $K_\a$, namely, Gaussian with mean $K$ and variance $\s_K^2$, one gets
\be
R(k)=\f{\text{max} (0, k - \mu_c\g^{-1}\bra \l\ket + z_\lambda \sqrt{ \s_c^2\g^{-1} q_\l})}{1-\g^{-1}\s_c^2\nu}.
\ee

Therefore, we compute
\bea
\bra kR(k)\ket_{z_\lambda, k}&=& \f{1}{\sqrt{2\pi}}\left\bra \int dk\, k R(k)e^{-\f{(k-K)^2}{2\s_K^2}}\right\ket_{z_\lambda}\\
&=&\frac{1}{1-\gamma^{-1}\sigma_c^2 \nu}\frac{1}{\sqrt{2\pi\sigma_K}} \left< \int_{-\infty}^{\infty} dk k \, \text{max}\left[0, k -\mu_c\gamma^{-1}\left< \lambda
\right>+ \sqrt{\sigma_c^2\gamma^{-1}q_\lambda} z_\lambda \right]e^{-\frac{(k-K)^2}{2\sigma_K^2}}\right>_{z_\lambda}\\
&=&\frac{1}{1-\gamma^{-1}\sigma_c^2 \nu} \int_{-\infty}^{\infty} \int_{-\infty}^{\infty}\frac{dk dz_\lambda}{2\pi\sqrt{\sigma_K}} k \, \text{max}\left[0, k -\mu_c\gamma^{-1}\left< \lambda
\right>+\sqrt{\sigma_c^2\gamma^{-1}q_\lambda} z_\lambda   \right]e^{-\frac{(k-K)^2}{2\sigma_K^2}}e^{-\frac{z_\lambda^2}{2}}
\label{eq:rk}
\eea

To simplify the calculation, let us introduce another Gaussian variable $z_K$ with zero mean and unit variance. The integral part can now be written as:
\bea
&&\int_{-\infty}^{\infty} \int_{-\infty}^{\infty}\frac{dz_Kd z_\lambda}{2\pi}  e^{-\frac{z_K^2+z_\lambda^2}{2}}(K+\sigma_K z_K)   \, \text{max}\left[0, \!K \!+ \!\sigma_K z_K \!- \!\mu_c\gamma^{-1}\left< \lambda
\right>+\sqrt{\sigma_c^2\gamma^{-1}q_\lambda} z_\lambda  \right]\\
&=&\int_{-\infty}^{\infty} \int_{-\infty}^{\infty}\frac{dz_Kd z_\lambda}{2\pi}  e^{-\frac{z_K^2+z_\lambda^2}{2}}K   \, \text{max}\left[0, \!K \!- \!\mu_c\gamma^{-1}\left< \lambda
\right>+\sigma_K z_K +\sqrt{\sigma_c^2\gamma^{-1}q_\lambda} z_\lambda    \right] \nonumber\\
&&+\int_{-\infty}^{\infty} \int_{-\infty}^{\infty}e^{-\frac{z_K^2+z_\lambda^2}{2}}\frac{dz_Kd z_\lambda}{2\pi} \sigma_K z_K \, \text{max}\left[0, \!K \!\!- \!\mu_c\gamma^{-1}\left< \lambda
\right>+\sigma_K z_K +\sqrt{\sigma_c^2\gamma^{-1}q_\lambda} z_\lambda  \right] \label{intcom}
\eea

Using $z_R \sqrt{\sigma^2_K+\sigma_c^2\gamma^{-1}q_\lambda}= \sigma_K z_K +\sqrt{\sigma_c^2\gamma^{-1}q_\lambda} z_\lambda $ , the first term of Equation (\ref{intcom}) can be written as 
 \bea
\int_{-\infty}^{\infty}\frac{d z_R}{\sqrt{2\pi}}  e^{-\frac{z_R^2}{2}}K   \, \text{max}\left[0, \!K \!- \!\mu_c\gamma^{-1}\left< \lambda 
\right>+z_R \sqrt{\sigma^2_K+\sigma_c^2\gamma^{-1}q_\lambda}   \right] = \sqrt{\sigma^2_K+\sigma_c^2\gamma^{-1}q_\lambda} K  w_1(\Delta),
\label{eq:int1}
\eea
where
\bea
\label{eq:Delta}
\Delta = \f{K-\mu_c\g^{-1}\bra \l\ket}{\sqrt{\sigma^2_K+\sigma_c^2\gamma^{-1}q_\lambda}}. 
\eea
 
Using integration by parts in the $z_K$ integral, we find that the second term of Equation (\ref{intcom}) is 
\bea
&&\int_{-\infty}^{\infty} \int_{-\infty}^{\infty}e^{-\frac{z_K^2+z_\lambda^2}{2}}\frac{dz_Kd z_\lambda}{2\pi} \sigma_K z_K \, \text{max}\left[0, \!K \!+ \!\sigma_K z_K \!- \!\mu_c\gamma^{-1}\left< \lambda
\right>+\sqrt{\sigma_c^2\gamma^{-1}q_\lambda} z_\lambda  \right]\nonumber \\
&=&\sigma_K^2 \int_{-\infty}^{\infty} \int_{-\infty}^{\infty}e^{-\frac{z_K^2+z_\lambda^2}{2}}\frac{dz_Kd z_\lambda}{2\pi}\Theta\left(\!K \!+ \!\sigma_K z_K \!- \!\mu_c\gamma^{-1}\left< \lambda
\right>+\sqrt{\sigma_c^2\gamma^{-1}q_\lambda} z_\lambda\right)
\eea
where $\Theta(x)$ equals 0 for $x<0$, and equals 1 for $x \geq 0$. It arises from taking the derivative of $\text{max}\left[0, \!K \!+ \!\sigma_K z_K \!- \!\mu_c\gamma^{-1}\left< \lambda
\right>+\sqrt{\sigma_c^2\gamma^{-1}q_\lambda} z_\lambda  \right]$ with respect to $z_K$ in the integration by parts. As in the first integral, we can now change variables to $z_R$, and use the $\Theta$ function to set the lower limit of integration:
\bea
&&\sigma_K^2 \int_{-\infty}^{\infty} \int_{-\infty}^{\infty}e^{-\frac{z_K^2+z_\lambda^2}{2}}\frac{dz_Kd z_\lambda}{2\pi}\Theta\left(\!K \!+ \!\sigma_K z_K \!- \!\mu_c\gamma^{-1}\left< \lambda
\right>+\sqrt{\sigma_c^2\gamma^{-1}q_\lambda} z_\lambda\right)\nonumber\\
&=& \sigma_K^2  \int_{-\Delta}^{\infty}e^{-\frac{z_R^2}{2}}\frac{dz_R}{\sqrt{2\pi}}\\
&=&\sigma_K^2 w_0(\Delta)
\label{eq:int2}
\eea
where $\Delta$ is the same quantity defined in Equation (\ref{eq:Delta}) above.

Putting Equations (\ref{eq:int1}) and (\ref{eq:int2}) back into Equation (\ref{eq:rk}), we finally find:
\bea
\bra kR(k)\ket_{z_\lambda, k}&=&\frac{1}{1-\gamma^{-1}\sigma_c^2 \nu}\left[\sigma^2_K w_0(\Delta) +\sqrt{\sigma^2_K+\sigma_c^2\gamma^{-1}q_\lambda} K w_1(\Delta)  \right].
\eea
\end{document}